\documentstyle[aps,epsfig]{revtex}
\begin{document}
\renewcommand{\thefootnote}{\dagger}
\def \be {\begin{equation}}
\def \ee {\end{equation}}
\def \bea {\begin{eqnarray}}
\def \eea {\end{eqnarray}}
\def \beax {\begin{eqnarray*}}
\def \eeax {\end{eqnarray*}}
\preprint{SNUTP 99-014; 
          KIAS-P99013}
\title{Synchronization in a System of Globally Coupled Oscillators\\
with Time Delay}
\author{M.~Y. Choi, H.~J. Kim, and D. Kim}
\address{Department of Physics and Center for Theoretical Physics\\
     Seoul National University,
     Seoul 151-742, Korea}
\author{H. Hong}
\address{Department of Physics Education and Center for Theoretical Physics\\
     Seoul National University,
     Seoul 151-742, Korea}   
\maketitle
\draft

\begin{abstract}
We study the synchronization phenomena in a system of globally
coupled oscillators with time delay in the coupling.
The self-consistency equations for the order parameter are derived,
which depend explicitly on the amount of delay.
Analysis of these equations reveals that the system in general exhibits
discontinuous transitions in addition to the usual continuous transition, 
between the incoherent state and a multitude of coherent states with different 
synchronization frequencies. 
In particular, the phase diagram is obtained on the plane of the coupling
strength and the delay time, and ubiquity of multistability as well as
suppression of the synchronization frequency is manifested.
Numerical simulations are also performed to give consistent results.
\end{abstract}



\section{Introduction}

When a large population of limit cycle oscillators with slightly different 
natural frequencies are coupled, they often come to oscillate with an identical frequency.
Such collective synchronization phenomena have been observed in various oscillatory systems 
in physics, biology, chemistry, and other sciences~\cite{Winfree,Walker,Eckhorn,Benz},
attracting much interest in recent years~\cite{Kura,Daido,Stro,Park,Choi,Arenas,Tanaka}. 
In the Kuramoto model for those oscillator systems,
oscillators are coupled with each other via the interaction 
which depends on the phase difference between each pair~\cite{Kura}.
It describes the emergence of
phase coherence with the increase of the coupling strength,
elucidating interesting connection between the collective synchronization and a phase 
transition.

Here, as in the usual dynamics of many-particle systems, sufficient attention has not been
paid to the effects of time retardation in the oscillator system.
In biological systems such as pacemaker cells and neurons, however, temporal delay is natural and
the finite time interval required for the information transmission 
between two elements may be important~\cite{Walker,Eckhorn}.
Time delay in the interaction may modify drastically dynamic behavior of the system,
such as stability and ergodicity~\cite{Choi85}.
In some types of a system of coupled oscillators, retarded 
interactions have been found to result in multistability and suppression of 
the collective frequency~\cite{Shuster,Luzyanina,Niebur,Nakamura}.
In a system of two coupled oscillators, it has been found that
the time delay induces a multitude of synchronized solutions.  Namely,
in the system with finite time delay, more than one 
stable solution are possible at given coupling strength. 
Among those, the most stable solution is the one with the largest 
synchronization frequency, as shown via the linear stability analysis~\cite{Shuster}.
Similar behaviors including frequency suppression and multistability
have been observed in the neural network model, where peripheral oscillators 
with identical natural frequencies are coupled only with a central oscillator 
by forward and backward connections with time delay~\cite{Luzyanina}. 
The two-dimensional system of identical 
oscillators with time-delayed nearest-neighbor coupling
has also been considered to reveal similar frequency suppression~\cite{Niebur}. 
The system of non-identical oscillators with delayed interactions has 
been studied recently~\cite{YS}: In the case of a Lorenzian
distribution of natural frequencies with a nonzero mean, 
the stability boundary of the incoherent state has been obtained and
coexistence of one or more coherent and/or incoherent states has been
observed in appropriate regions~\cite{YS}.
However, detailed behaviors such as frequency suppression and emergence
of different coherent states have not been addressed fully.

This paper investigates in detail the effects of time delay in the
interaction on collective synchronization of coupled oscillators
with different natural frequencies.
For this purpose, we derive the self-consistency equations for the
order parameter and examine how the characteristic features of the collective
synchronization change due to time delay.
This reveals a multitude of coherent states with nonzero
synchronization frequencies, each separated from the
incoherent state by a discontinuous transition.
In particular, we show that the system with a nonzero average frequency
can be reduced to the system with the vanishing average natural frequency,
which allows us to focus on the latter system.
As in the system without delay, there exists the critical coupling strength,
at which the system undergoes the usual continuous transition from the incoherent
state to the coherent state displaying collective synchronization 
(with zero synchronization frequency).
In addition, at higher values of the coupling strength (beyond the critical value), 
coherent states with larger
synchronization frequencies also appear via discontinuous transitions.
Thus coherent states with different synchronization frequencies
in general coexist in the appropriate regions, leading to multistability.
The synchronization frequency of the oscillators in a coherent state
is observed to decrease with the delay time,
which is similar to the result of other systems with time
delay~\cite{Luzyanina,Niebur,Nakamura}.

There are five sections in this paper:
Section II presents the system of globally coupled oscillators with time 
delay, as a generalization of the Kuramoto model.
The stationary probability distribution for the system is obtained, and
the self-consistency equations for the order parameter are derived.
Section III is devoted to the analysis of the self-consistency equations,
which reveals the characteristic behavior of the system as the coupling
strength or the delay time is varied.
In particular, the phase diagram is obtained on the plane of the coupling
strength and the delay time, and
ubiquity of multistability as well as suppression of the synchronization
frequency is demonstrated.
Numerical simulations are also performed and the results,
which are in general consistent with the analytical ones, are presented 
in Sec. IV.  
Finally, Sec. V summarizes the main results, while
some details of the calculations are presented in Appendices A and B.
    
\section{System of coupled oscillators with time delay}

The set of equations of motion for $N$ coupled oscillators, each 
described by its phase $\phi_i~ (i=1,2,\cdots,N)$, is given by
\begin{equation} \label{model}
\dot\phi_i (t) =\omega_i - \frac{K}{N} {\sum_j}^{'}
\sin[\phi_i(t)-\phi_j(t{-}\tau)],
\end{equation}
where the prime restricts the summation such that $j\neq i$.
The first term on the right-hand side represents the natural frequency
of the $i$th oscillator, which is distributed 
according to the distribution function $g(\omega)$.
Here $g(\omega)$ is assumed to be smooth and symmetric about $\omega_0$,
which may be taken to be zero without loss of generality (see below), and also
to be concave at $\omega=0$, i.e., $g''(0) < 0 $.
The second term denotes the global coupling of strength $K/N$ 
between oscillators, with time delay, indicating that each oscillator 
interacts with other oscillators only after the retardation time $\tau$.
Without time delay, Eq.~(\ref{model}) exactly reduces to the Kuramoto model.

In order to describe collective synchronization of such an $N$ oscillator 
system, we define the complex order parameter, whose amplitude
represents the degree of synchronization, to be 
\begin{equation} \label{order}
\Psi\equiv \frac{1}{N}\sum_{j=1}^{N} e^{i \phi_j} =\Delta e^{i \theta}.
\end{equation}
Here it is convenient to introduce new variables $\psi_i$ 
defined by $\psi_{i} \equiv \phi_i-\Omega t$, where $\Omega $ is a constant.
Note the existence of physical invariance due to the rotational symmetry of the 
total system.
In terms of the new variables, Eq.~(\ref{model}) reads 
\begin{equation} \label{newmodel}
\dot\psi_i =\tilde{\omega_i}-\frac{K}{N} {\sum_j}^{'}
\sin[\psi_i(t)-\psi_j(t{-}\tau)+\Omega \tau],
\end{equation}
where $\tilde{\omega_i} \equiv \omega_i-\Omega $.
Multiplying Eq.~(\ref{order}) by $e^{-i \Omega t}$, we also obtain the 
corresponding order parameter for the new variables
\begin{equation} \label{neworder}
\tilde{\Psi} \equiv \frac{1}{N}\sum_{j=1}^{N} e^{i \psi_j} =\Delta e^{i
\tilde{\theta}},
\end{equation}
where $\tilde{\theta} \equiv \theta - \Omega t$.
Incidentally, the order parameter defined in Eq.~(\ref{neworder}) allows us 
to reduce Eq.~(\ref{newmodel}) into a single decoupled equation with time delay
\begin{equation} \label{model1}
\dot\psi_i =\tilde{\omega_i} - K \Delta \sin(\psi_i-\theta_0),
\end{equation}
where $\theta_0 \equiv \tilde{\theta}-\Omega \tau$.
Although $\Delta$,~$\Omega$, and ~$\theta_{0}$ depend on the delay time $\tau$,
they are assumed to be independent of time $t$, which is possible due to the symmetry.
Considering the relation between the old order parameter and the new one
\be
\Psi = \tilde{\Psi} e^{i \Omega t},
\ee
we understand that the collective synchronization can be described 
in terms of a giant oscillator rotating with the frequency $\Omega$
which is in general nonzero.
For finite delay time, there exists a multitude 
of synchronized solutions with nonzero values of $\Omega$; this is in contrast 
to system without delay, where the rotational symmetry of the system allows
us to set $\Omega = 0$.

Instead of Eq.~(\ref{model1}), which may be regarded as a Langevin equation without noise, 
one may resort to the corresponding Fokker-Planck equation for the probability distribution
$P(\psi,t)$ at zero temperature~\cite{Risken89}:
\begin{equation} \label{fokker}
\frac{\partial}{\partial t}P(\psi,t)
= \frac{\partial}{\partial \psi}
[K \Delta \sin (\psi-\theta_0)-\tilde{\omega}]P(\psi,t).
\end{equation}
The order parameter given by Eq.~(\ref{neworder}) then obtains the form
\begin{eqnarray} \label{self}
\Delta e^{i \tilde{\theta}}
 &=&  \frac{1}{N}\sum_{j=1}^{N} e^{i \psi_j}  \nonumber\\
 &=& \int_{-\infty}^{\infty} ~d\tilde{\omega} ~g(\tilde{\omega}{+}\Omega)
    ~\langle e^{i\psi}\rangle_{t-\tau; \tilde{\omega}},
\end{eqnarray}
where $\tilde{\omega} =\omega-\Omega$ has been noted, and
the average of $e^{i \psi}$ is to be taken over the distribution
$P(\psi, t{-}\tau)$ of Eq.~(\ref{fokker}) with given $\tilde{\omega}$:
$
\langle e^{i \psi}\rangle_{t-\tau; \tilde{\omega}}
 \equiv \int_{0}^{2 \pi} d\psi \,P(\psi, t{-}\tau) \,e^{i\psi} .
$

In the stationary state, we take the average over the stationary distribution
$P^{(0)}(\psi;\tilde{\omega})$ of Eq.~(\ref{fokker}).
With the stationary solution 
\begin{equation} \label{statsol}
P^{(0)}(\psi;\tilde{\omega})
= \left\{
        \begin{array}{ll}
         \delta\left[ \psi-\theta_{0}-\sin^{-1}(\tilde{\omega}/K \Delta)\right] &{\rm for}
               \quad | \omega | \leq  K \Delta
                         \\
        \displaystyle
         \frac{\sqrt{\tilde{\omega}^2-(K \Delta)^2}}{
          2\pi \left| \tilde{\omega}-K \Delta\sin(\psi-\theta_{0})
               \right|} &{\rm otherwise},
        \end{array}
  \right.
\end{equation}
it is easy to compute the average:
\begin{eqnarray} \label{avstat}
\langle e^{i \psi}\rangle_{\tilde{\omega}}
& \equiv & \int_{0}^{2 \pi} d\psi \,P^{(0)}(\psi;\tilde{\omega})~e^{i\psi} \nonumber \\
& = & e^{i\theta_0} \left\{\begin{array}{ll}
     i(\tilde{\omega}/K\Delta)
       - i \sqrt{(\tilde{\omega}/K\Delta)^2-1}, & \mbox{~~~ $\tilde{\omega}> K \Delta$} \\
     i(\tilde{\omega}/K\Delta)
        + \sqrt{1-(\tilde{\omega}/K\Delta)^2}, &  \mbox{~~~ $ -K\Delta\leq \tilde{\omega}\leq K\Delta$}\\
     i(\tilde{\omega}/K\Delta)
       + i \sqrt{(\tilde{\omega}/K\Delta)^2-1}, & \mbox{~~~ $\tilde{\omega}< -K\Delta$}.
  \end{array}
  \right.
\end{eqnarray}

It is thus natural to divide the system into two groups: one satisfying
$|\tilde{\omega}| \leq K\Delta$, which is called the synchronization group, 
and the other $|\tilde{\omega}| > K\Delta$, the desynchronization group.
Accordingly, we write
\begin{equation} \label{magorder}
\Delta = \Delta_{s} + \Delta_{d},
\end{equation}
where $\Delta_{s/d}$ is the contribution from the synchronization/desynchronization 
group to the order parameter.
The contribution from the synchronization group is given by
\begin{equation} \label{synself1}
\Delta_{s} \,e^{i \Omega\tau}
= K\Delta \int_{-1}^{1} dx\, g(\Omega{+}K\Delta x)\,[\sqrt{1-x^{2}}+ix],
\end{equation}
where $x \equiv \tilde{\omega}/K\Delta$.
Separating Eq.~(\ref{synself1}) into the real and imaginary parts,
we obtain the two coupled nonlinear equations
\begin{eqnarray} \label{synself2}
\Delta_{s}\cos \Omega\tau
&=& K\Delta \int_{-1}^{1} dx\,g(\Omega{+}K\Delta x)\,\sqrt{1-x^{2}},\nonumber\\
\Delta_{s}\sin \Omega\tau
&=& K\Delta\int_{-1}^{1} dx\,g(\Omega{+}K\Delta x)\,x.
\end{eqnarray}
Similarly, the desynchronization group leads to the equation 
\begin{eqnarray} \label{desynself1}
\Delta_{d}\, e^{i\Omega\tau }
=&& iK\Delta \int_{1}^{\infty} dx \,g(\Omega{+}K\Delta x) \,\left(x-\sqrt{x^{2}-1}\right) \nonumber \\
 && +iK\Delta \int_{-\infty}^{-1} dx \,g(\Omega{+}K\Delta x) \, \left(x+\sqrt{x^{2}-1}\right)
\end{eqnarray}
or
\begin{eqnarray} \label{desynself2}
\Delta_{d}\cos \Omega\tau &=& 0, \nonumber\\
\Delta_{d} \sin \Omega\tau
&=& K\Delta \left[ \int_{1}^{\infty} dx \,g(\Omega{+}K\Delta x) \,\left(x-\sqrt{x^{2}-1}\right)
            \right. \nonumber \\
&&~~~~~\left.+\int_{-\infty}^{-1} dx \, g(\Omega{+}K\Delta x) \, \left(x+\sqrt{x^{2}-1}\right)\right].
\end{eqnarray}

Note that, unlike the Kuramoto model, the imaginary parts of 
Eqs.~(\ref{synself1}) and (\ref{desynself1}) do not vanish,
which arises from the fact that 
the nonzero collective frequency due to time delay breaks the symmetry 
of the integration interval of the distribution $g(\omega)$.
It is obvious in Eq.~(\ref{desynself2}) that 
the contribution from the desynchronization group vanishes in the absence of time delay
($\tau =0$).
Recalling that in the presence of time delay 
the total order parameter is given by the sum of the two contributions, 
one from the synchronization group and the other from the desynchronization group,
we finally obtain the self-consistency equations
from Eqs.~(\ref{synself2}) and (\ref{desynself2}):
\begin{eqnarray} \label{self2}
\Delta\cos \Omega\tau       
&=& K\Delta \int_{-1}^{1} dx \,g(\Omega{+}K\Delta x) \,\sqrt{1-x^{2}}, \nonumber\\
\Delta\sin \Omega\tau
&=& K\Delta \left[\int_{-1}^{1} dx \,g(\Omega{+}K\Delta x)\,x 
                + \int_{1}^{\infty} dx \,g(\Omega{+}K\Delta x)\, \left(x-\sqrt{x^{2}-1}\right)
            \right.  \nonumber\\
        &&~~~~~~~\left.~+ \int_{-\infty}^{-1} dx \,g(\Omega{+}K\Delta x) \,\left(x+\sqrt{x^{2}-1}\right)\right].
\end{eqnarray}

When the average natural frequency is not zero ($\omega_0 \neq 0$),
we define the variables $\psi_i \equiv \phi_i - (\Omega + \omega_0 )t$, and obtain exactly Eq.~(\ref{self2})
except for $\Omega$ replaced by $\tilde{\Omega} \equiv \Omega + \omega_0$.
For example, the first equation is given by
\begin{equation} \label{unsym}
\Delta\cos \tilde{\Omega}\tilde{\tau}
 = K\Delta \int_{-1}^{1} dx \,\tilde{g}(\tilde{\Omega}{+}K\Delta x) \,\sqrt{1-x^{2}},
\end{equation}
where $\tilde{\tau}$ is the delay time and the distribution 
$\tilde{g}(\omega)$ is symmetric about $\omega=\omega_0$.
Since the distribution $g(\omega) \equiv \tilde{g}(\omega{+}\omega_0)$ is symmetric 
about $\omega =0$, we rewrite the above equation in the form
\begin{equation} \label{unsym2}
\Delta\cos [(\Omega{+}\omega_0)\tilde{\tau}]
 = K\Delta \int_{-1}^{1} dx \, g(\Omega{+}K\Delta x) \,\sqrt{1-x^{2}},
\end{equation}
which, with the identification $(\Omega+\omega_0)\tilde{\tau}\equiv \Omega\tau$, 
just reproduces the first equation in Eq.~(\ref{self2}).
Accordingly, the behavior of the system with $\omega_0 \neq 0$,
which has been considered in Ref.~\cite{YS}, can be obtained from
that of the system with $\omega_0=0$ via appropriate rescaling of parameters.

\section{Analysis of the self-consistency equations}

The non-vanishing imaginary part of the self-consistency equation given by 
Eq.~(\ref{self2}), arising from time delay, leads to  
a variety of behaviors which are not displayed by the system without delay.
In this section, we solve the two coupled equations in Eq.~(\ref{self2}) to obtain 
the synchronization frequency $\Omega$ and the order parameter $\Delta$.
We take the Gaussian distribution with zero mean and 
unit variance for the natural frequencies:
$g(\omega) = (2\pi)^{-1/2} e^{-\omega^2/2}$,
and first compute numerically the synchronization frequency and the order parameter
for various values of the coupling strength $K$ and the delay time $\tau$. 

Figure~1 exhibits the dependence of the synchronization frequency on $K$ and
$\tau$, which manifests multistability. 
At small values of $\tau$, only the nontrivial solution $(\Delta\neq 0)$ with
$\Omega=0$ appears for $K>K_c \,(\approx 1.596)$, as in the system without delay.
For large $\tau$, on the other hand, solutions with $\Omega\neq 0$
also emerge as $K$ is increased further.
In Fig.~1(a), where the synchronization frequency is plotted as a function of 
the delay time $\tau$ at $K = 10$, it is also observed that the synchronization 
frequency is suppressed as time delay is increased. This is expected since the 
delay tends to disturb synchronization~\cite{Luzyanina}.
In Fig.~1(b), we plot the synchronization frequency as a function of the
coupling strength at $\tau = 5$.
It shows that at given values of $\tau$ the synchronization frequency 
$\Omega$ depends rather weakly on $K$ after synchronization sets in.
Among those solutions at given coupling strength $K$, the most stable solution is the one 
with the largest value of $\Omega$~\cite{Shuster} although the basin of
attraction in general shrinks with $\Omega$.

The phase boundaries separating the coherent states ($\Delta\neq 0$) 
with various synchronization frequencies from the incoherent state ($\Delta=0$)
are shown in Fig.~2, where data have been taken with the step width $\delta\tau =0.06$.
Below the lowest boundary, which is the straight line $K=K_c \approx 1.596$,
only the incoherent state exists; above it, the coherent state with $\Omega=0$ also exist.
Similarly, above each boundary in Fig.~2(a), 
a new additional coherent state with a larger synchronization frequency emerges.
Note here that the region of the existence of coherent states constitutes 
two-dimensional (semi-infinite) surfaces in the three-dimensional $(K, \tau, \Omega)$ space.
Whereas Figs.~1(a) and (b) may be regarded as the cross-sections of these
surfaces at given values of $K$ and of $\tau$, respectively, Fig.~2(a)
represents the projection of the {\em boundaries} of these surfaces onto the $K$-$\tau$ plane.
In Fig.~2(b) the curves of Fig.~2(a) are redrawn, with the horizontal axis rescaled:
$\tilde{\tau}$ in (b) corresponds to $\Omega\tau/(\Omega+3)$ in (a).  
In this new scale of the horizontal axis, unlike in Fig.~2(a),
the boundaries intersect with each other, and only the envelope consisting
of the curve segments with lowest values of $K$ at given $\tilde{\tau}$ 
is displayed in Fig.~2(b).
According to the discussion in Sec.~II, Fig.~2(b) describes the phase boundary
below which only the incoherent state exists in the system with $\omega_0 =3$.
Above the boundary, 
the coherent state with the appropriate (nonzero) synchronization frequency,
depending on $\tilde{\tau}$,
appears and can coexist with the incoherent state.
Note that the lowest boundary ($K=K_c \approx 1.596$) in Fig.~2(a) 
has no counterpart in Fig.~2(b) since the zero synchronization frequency ($\Omega = 0$)
corresponds to $\tilde{\tau}=0$.
Similar boundaries have been obtained through numerical simulations
for the Lorenzian as well as the delta-function distributions~\cite{YS}.
It is of interest to compare Fig. 2(b) with Fig. 4 in Ref.~\cite{YS},
which indicates that the Gaussian distribution leads to smoother stability boundaries 
than the Lorenzian distribution.
%

In Fig.~3, the obtained order parameter $\Delta$ is depicted as a function
of $K$ at $\tau = 5$. Each line describes the transition for each synchronization
frequency, and the critical coupling strength $K_c$ is shown to increase for the transition 
with larger $\Omega$. For example, the leftmost curve, which corresponds to $\Omega = 0$,
gives $K_c \approx 1.596$, whereas the next one, corresponding to $\Omega
\approx 1.09$, gives $K_c \approx 1.97$. 
Note that as $K$ approaches $K_c\,(\approx 1.596)$, the order parameter
with $\Omega=0$ decreases continuously to zero, indicating that 
the leftmost curve describes a continuous transition at $K_c$.
On the other hand, the rest of the curves with $\Omega > 0$
apparently display jumps in the order parameter, 
indicating discontinuous transitions.
Accordingly, whereas the lowest boundary in Fig.~2(a) describes a continuous transition,
the others as well as the boundary in Fig.~2(b) correspond to discontinuous transitions.

To understand the nature of these transitions analytically, we
assume $K\Delta\ll 1$ near the transition to the coherent state
and expand Eq.~(\ref{self2}) to the order of $(K\Delta)^3$, 
together with $\Omega$ also expanded accordingly:
\be    \label{ao}
\Omega\approx \Omega_{0}+\Omega_{1}K\Delta+\Omega_{2}{(K\Delta)}^2.
\ee
We investigate two regimes, $\Omega \ll K\Delta$ and  
$\Omega \gg K\Delta$, which exhibit 
phase transitions of different types with each other,
still taking the Gaussian distribution with zero mean and 
unit variance for $g(\omega)$. 

For $\Omega \ll K\Delta \,(\ll 1)$,
the self-consistency equation for the order parameter 
takes the form
\be    \label{asc1}
\Delta = a_1 K\Delta+b_1 (K\Delta)^2+c_1 (K\Delta)^3+{\cal O}(K\Delta)^4,
\ee
where the coefficients $a_1$, $b_1$, and $c_1$ depend on $\Omega_0$, $\Omega_1$, and $\Omega_2$
defined in Eq.~(\ref{ao}). 
Their specific forms as well as the details of the calculation are given 
in Appendix A.
Since the condition $\Omega \ll K\Delta$ implies $\Omega_0 \ll 1$, we need to 
consider Eq.~(\ref{omega1}) only for the range
$-\pi/2 < \tan^{-1} \left(\sqrt{2/\pi}\Omega_0 e^{\Omega_0^2/2}\right) < \pi/2$.
It is then obvious that the desired solution of Eq.~(\ref{omega1}) is simply
$\Omega_0 = \Omega_1 = \Omega_2 = 0$, regardless of $\tau$.
Inserting these values into Eq.~(\ref{cof1}), we obtain the values of the coefficients:
$a_1 = 0.626$, $b_1 = 0$, and $c_1 = -0.078$. 
Figure~4 illustrates the graphical solution of Eq.~(\ref{asc1}), displaying
$f(\Delta)\equiv (a_1 K-1)\Delta+b_1 (K\Delta)^2+c_1 (K\Delta)^3$ versus $\Delta$
for $b_1 =0$.
Note that the critical coupling strength is given by $K_c\,(\equiv {a_1}^{-1}) 
\approx 1.595$, which indeed agrees perfectly with the numerical value given by the
leftmost curve in Fig.~3.
It is thus concluded that the system displays a continuous phase transition with $\Omega = 0$, 
which is consistent with the result of the Kuramoto model. 

In the opposite case of $\Omega \gg K\Delta \,(\ll 1)$,
we can still obtain the self-consistency equation for $\Delta$
up to the order of $(K\Delta)^3$, in a manner similar to that for the 
previous small-$\Omega$ case: 
\be    \label{asc2}
\Delta = a_2 K\Delta+b_2 (K\Delta)^2+c_2 (K\Delta)^3+{\cal O}(K\Delta)^4,
\ee
where the coefficients $a_2$, $b_2$, and $c_2$ again depend on $\Omega_0$, $\Omega_1$, and 
$\Omega_2$ (see Appendix B).
In this case, we need to obtain larger solutions, considering the regions
$(n+1/2)\pi < \tan^{-1} \left(\sqrt{2/\pi}\Omega_0 e^{\Omega_0^2/2}\right) < (n+3/2)\pi$ 
with non-negative integer $n$. 
Interestingly, this in general yields nonzero values of $\Omega_0$ and accordingly,
nonzero values of $b_2$, with which Eq.~(\ref{asc2}) displays a jump in $\Delta$
at $K =  -4c_2({b_2}^2-4c_2a_2)^{-1}$, thus indicating a discontinuous
transition~\cite{Choi}.
Such discontinuous transitions are ubiquitous for regions with
higher values of $n$.  Namely, the system with delay is in general characterized
by nonzero values of the synchronization frequency together with discontinuous
transitions, which is consistent with the numerical results displayed in Fig.~3.
There the jumps in $\Delta$ displayed by the curves with $\Omega > 0$,
associated with discontinuous transitions,
may invalidate the assumption $K\Delta \ll 1$, and
the expansion in Eq.~(\ref{asc2}) is not expected to yield
quantitatively accurate results.
Nevertheless the appearance of such discontinuous transitions has been revealed 
by the above expansion, which is concluded to give a qualitatively correct description 
of the nature of transitions.

For more accurate results,
we investigate these transition phenomena by examining in detail the behaviors
of the solutions of the self-consistency equations.
Note that $\Delta =0$ is always a solution of Eq.~(\ref{self2}) for all values of $\Omega$.
To seek for other solutions, we divide Eq.~(\ref{self2}) by $\Delta$
and obtain
\begin{eqnarray} \label{self20}
K^{-1}\cos \Omega\tau 
&=& \int_{-1}^{1} dx \,g(\Omega{+}K\Delta x) \,\sqrt{1-x^{2}}, \nonumber\\
K^{-1}\sin \Omega\tau
&=& \int_{-\infty}^{\infty} dx \,g(\Omega{+}K\Delta x)\,x 
                - \int_{1}^{\infty} dx \,g(\Omega{+}K\Delta x)\, \sqrt{x^{2}-1} \nonumber \\
  & &~~ + \int_{-\infty}^{-1} dx \,g(\Omega{+}K\Delta x)\,\sqrt{x^{2}-1},
\end{eqnarray}
which may be computed numerically.
The resulting values of $\Omega$ versus $\Delta$
are plotted in Figs.~5-7 for $\tau=5$. 
The solid and the dotted lines represent solutions
of the first (real part) and the second (imaginary part) equations 
of Eq.~(\ref{self20}), respectively.  
In each figure, the point where the two lines meet 
with each other provides the 
synchronization frequency $\Omega$ and the order parameter $\Delta$.  
Figure~5(a) shows the absence of a meeting point for $K=1.58$,
which implies that synchronization does not set in yet.
In contrast, the meeting of the solid and dotted lines is obvious for $K = 1.60$ shown in (c); 
(b) reveals a continuous transition (for $\Omega = 0$) at 
$K = K_c \approx 1.596$, which coincides with the previous result.
The value of $\Delta$ grows continuously as $K$ is increased beyond $K_c$ (see Fig.~6).
When $K$ reaches the value $1.97$, as displayed in Fig.~6(b), 
there emerges via a tangent bifurcation
an additional meeting point at finite values of $\Delta \,(\approx 0.08)$
and $\Omega \,(\approx 1.09)$,
giving rise to a discontinuous transition, in agreement with the result shown in Fig.~3.
As the coupling strength is increased further, there appear two meeting points, 
giving two values of the order parameter for the pair of the lines
(i.e., with almost the same value of $\Omega$), as shown in Fig.~6(c).
Such a tangent bifurcation in general produces a pair of stable
and unstable solutions; 
here the solution with the smaller value of $\Delta$, decreasing with $K$,
should be unstable.
Figure~7(c) shows that the unstable solution becomes null ($\Delta =0$)
at $\Omega \approx 1.257$ as $K$ approaches $K_0 \approx 3.515$.
Figure~7 also reveals the occurrence of the third transition 
at $K_c\approx 3.46$, which is of the same nature as the second.

The values of $K_c \approx 1.596$ and $K_0 \approx 3.515$ can also be obtained
analytically since they are given by
the solutions of Eq.~(\ref{self20}) in the limit $\Delta\rightarrow 0$.
In this limit, the right-hand side of the second equation vanishes, 
yielding $\Omega = n\pi/\tau$ with $n$ integer.  
The first one, which reduces to $K^{-1}\cos\Omega\tau = (\pi/2) g(\Omega)$,
then gives
\begin{equation} \label{K}
K = \frac{2}{\pi g(2n\pi/\tau)},
\end{equation}
where it has been noted that $K>0$. 
Taking $n=0$ and $n=1$ in Eq.~(\ref{K}), where $g(\omega)$ is given by
the Gaussian distribution with unit variance, 
we obtain $K =K_c = \sqrt{8/\pi}\approx 1.596$ (with $\Omega=0$)
and $K=K_0 = \sqrt{8/\pi} e^{2(\pi/\tau)^2} \approx 3.515$ 
(with $\Omega =2\pi/\tau \approx 1.257$) for $\tau=5$, respectively.

To examine how the stability changes at these bifurcations,
we now consider a small perturbation from the incoherent state, for which
the stationary distribution in Eq.~(\ref{statsol}) is simply given by 
$1/2\pi$, and write
\begin{equation}  \label{perturb}
P(\psi, t) = \frac{1}{2 \pi} + \epsilon \eta(\psi, t),
\end{equation}
where $\epsilon \ll 1$.
Upon substitution into Eq.~(\ref{fokker}) and with 
$\Delta = \Delta_1 \epsilon + {\cal O}(\epsilon^2)$,
we obtain, to the lowest order in $\epsilon$,
\begin{equation} \label{fokker1}
\frac{\partial \eta}{\partial t} 
=  - ~\tilde{\omega} \frac{\partial \eta}{\partial \psi} 
   + \frac{K}{2 \pi} \Delta_1 \cos(\psi - \theta_0),
\end{equation}
and seek solutions of the form
\begin{equation} \label{solution1}
\eta(\psi,t) 
= c(t; \tilde{\omega}) e^{i \psi} + c^{*}(t; \tilde{\omega}) e^{-i \psi} ,
\end{equation}
where higher harmonics have been neglected.
Equations~(\ref{fokker1}) and (\ref{solution1}), together with
Eq.~(\ref{self}), lead to the amplitude equation for $c(t; \tilde{\omega})$:
\begin{equation} \label{fokker2}
\frac{\partial c(t, \tilde{\omega})}{\partial t}  
= - i \tilde{\omega} ~c(t, \tilde{\omega}) + \frac{K}{2} e^{i \Omega \tau}
\int_{-\infty}^{\infty} d\tilde{\omega} g(\tilde{\omega}+\Omega) c(t{-}\tau, \tilde{\omega}),
\end{equation}
which in general possesses both discrete and continuous spectra.
To find out the discrete spectrum, we put
\begin{equation} \label{solution2}
c(t;\tilde{\omega})
= b(\tilde{\omega}) \,e^{\lambda t},
\end{equation}
where the eigenvalue $\lambda$ is independent of $\tilde{\omega}$, 
and obtain the equation
\begin{equation} \label{fokker3}
e^{-(\lambda-i \Omega) \tau} ~\frac{K}{2} \int_{-\infty}^{\infty} 
d \omega ~\frac{g(\omega)}{\lambda + i ~(\omega - \Omega)} = 1 ,
\end{equation}
which has been examined for a Lorenzian distribution~\cite{YS}.

Here we investigate Eq.~(\ref{fokker3}) for a Gaussian distribution.
The stability of the incoherent state depends on whether all roots of
Eq.~(\ref{fokker3}) possess negative real parts, i.e., 
$\mbox{Re} \,\lambda < 0$.
This is the case for $K$ less than $K_s$, where the incoherent state 
is neutrally stable (since the continuous spectrum is pure imaginary).
Beyond $K_s$, there appears an eigenvalue with a positive real part,
giving rise to instability.
The value of $K_s$ can be computed from Eq.~(\ref{fokker3}) with
$\mbox{Re} \,\lambda = 0$ imposed;
this yields the coupled equations for $K_s$ and $\mbox{Im}\,\lambda$ 
\begin{eqnarray} \label{stab}
\cos \gamma\tau &=& K_s \sqrt{\frac{\pi}{8}}  e^{-\gamma^2/2}  \nonumber \\
\sin \gamma\tau &=& - \frac{K_s}{2} \gamma e^{-\gamma^2/2} 
                                           \sum_{k=0}^{\infty} \frac{\gamma^{2k}}{2^k (2k+1)k!},
\end{eqnarray}
where $\gamma \equiv \Omega -\mbox{Im}\,\lambda$.
Unlike the system without delay, Eq.~(\ref{stab}) has an infinite number of solutions,
among which the lowest value of $K_s$ should be taken.
It is obvious that $\gamma =0$ is the desired solution (regardless of time delay), leading to 
the lowest value $K_s = \sqrt{8/\pi} \approx 1.596$.
Note also that this value of $K_s$ coincides exactly with that of $K_c$,
implying that the incoherent state becomes unstable simultaneously
with the appearance of the (stable) coherent state with $\Omega =0$.

These results reveal that the order parameter exhibits a supercritical bifurcation 
at $K = K_c$ along the leftmost curve ($\Omega=0$) in Fig.~3.
Namely, the emergence of a nontrivial solution ($\Delta > 0$) 
is accompanied by the loss of stability of the null solution ($\Delta =0$) 
at $K_c(\Omega=0)\approx 1.596$.
For the rest ($\Omega >0$), on the other hand, 
the unstable solution, generated together with the stable one by a tangent bifurcation
at $K_c(\Omega)$,
decreases as $K$ is raised further and vanishes to zero at a larger value, 
$K=K_0 (\Omega)$.
For example, the unstable solution for $\Omega \approx 1.09$, emerging
at $K\approx 1.97$, decreases to zero at $K\approx 3.515$ [see Fig.~7(c)].
It is thus concluded that for $\Omega >0$ the bifurcation at $K_0$ is subcritical:
Between $K_c$ and $K_0$ there exist an unstable coherent state in addition to
the stable coherent states (and the incoherent one) although the unstable states
have not been displayed in Fig.~3.

The general features of the synchronization behavior obtained here
are similar to those in Ref.~\cite{YS}, and it
is thus concluded that the difference in the distribution of
natural frequencies does not change results qualitatively. 
On the other hand, we have examined additional interesting phenomena 
such as frequency suppression and details of multistability.
In particular, unlike in the system with $\omega_0 \neq 0$ considered mostly
in Ref.~\cite{YS}, here the phase boundaries with different values of
the synchronization frequency do not intersect with each other on the $K{-}\tau$ plane.
[Compare Fig.~2(a) and (b).]
Accordingly, the system with $\omega_0=0$ does not undergo a discontinuous
transition directly from the {\em stable} incoherent state and the coherent one with a nonzero
synchronization frequency, and the associated hysteresis may not be observed.
Further, in order to confirm these results, we have also performed numerical simulations,
the results of which are presented in the next section.

\section{Numerical simulations}

We have studied directly the equations of motion given by Eq.~(\ref{model}) via 
numerical simulations.  The globally coupled system of size $N=5000$,
where natural frequencies are distributed according to the 
Gaussian distribution with unit variance, has been considered, and
the Euler method with discrete time steps of $\delta t=0.01$ has 
been employed.  At each run, we have discarded the first $10^5$ time
steps per oscillator to eliminate transient effects and taken the next
$10^5$ time steps per oscillator to investigate synchronized solutions.
Finally, independent runs with 30 different realizations of the 
natural frequency distribution and initial conditions have been performed, 
over which the averages have been taken.  
In the simulations, the synchronization
frequency is given by the average phase speed, i.e., the average rate of 
the phase change, and the obtained data at the coupling strength $K=10$ 
are represented by crosses in Fig.~8(a).  Note that both the 
incoherent state and the coherent one are found to be stable at the same
value of $\tau$, indicating multistability; frequency suppression with 
increasing delay is also manifested.  For comparison, the results shown in 
Fig.~1(a), obtained from Eq.~(\ref{self2}), are also displayed, and perfect agreement
is observed.  Notice here that the basin of attraction shrinks rapidly
with the synchronization frequency $\Omega$, which makes it quite 
difficult in numerical simulations to find the coherent-state solutions 
with large values of $\Omega$.  Figure 8(b) shows the behavior of the
order parameter $\Delta$ as a function of the coupling strength $K$ for
$\tau=0$ (plus signs) and $\tau=5$ (crosses).  In both cases the system
displays a continuous transition to the coherent state (with zero 
synchronization frequency).  Slight suppression of synchronization by 
time delay can be observed.  The error bars have been estimated by the 
standard deviation and the lines are guides to the eye.
To make comparison of the analytical results obtained from Eq.~(\ref{self2}) and
the simulation results, we have also included in Fig.~8(b) the analytical results for $\tau=5$,
which are represented by the solid line.
Good overall agreement between the two can be observed.

\section{Summary}

We have studied analytically and numerically the collective
synchronization phenomena in a set of globally coupled oscillators with
time retarded interaction.
In order to understand the effects of time delay on the synchronization,
we have derived the self-consistency equations for the order parameter,
which describe synchronization in the system.
The detailed analysis of the self-consistency equations
has revealed a multitude of coherent states with nonzero synchronization frequencies, 
each separated from the incoherent state by a discontinuous transition.
At the critical coupling strength, the system exhibits the usual continuous transition 
from the incoherent state to the coherent one, displaying collective synchronization 
with zero synchronization frequency.
As the coupling strength is increased further,
coherent states with larger synchronization frequencies have also been shown to 
appear via discontinuous transitions from the incoherent state.
Thus a multitude of coherent states with different synchronization frequencies
have been found to coexist in the appropriate regions, leading to multistability.
The synchronization frequency of the oscillators in a coherent state
has been observed to decrease with the delay time.

To confirm the analytical results, we have also performed numerical simulations,
the results of which indeed
display multistability and suppression of the synchronization frequency. 
For detailed comparison, however, one should search the solution space extensively,
with varying initial conditions, to obtain solutions with various values of
the synchronization frequency.  This requires more extensive simulations, which
is left for future study.
Finally, one may also include stochastic noise in the system and study
its effects on synchronization behavior.  In particular, the interplay between the external
driving and noise poses the possibility of stochastic resonance~\cite{Hong99}, and it is of
interest to examine how the collective synchronization together with the time delay
affects the possible resonance phenomena.


\section*{Acknowledgments}
We thank G. S. Jeon, M.-S. Choi, and K. Park for illuminating discussions and 
C. W. Kim for the hospitality during our stay at Korea Institute for Advanced Study, where
part of this work was accomplished.
This work was supported in part by the SNU Research Fund, by the Korea Research Foundation, 
and by the Korea Science and Engineering Foundation.

\section*{Appendix A}
\renewcommand{\theequation}{A.\arabic{equation}}
\setcounter{equation}{0}

In the case $\Omega \ll K\Delta \,(\ll 1)$,
we approximate the integral appearing in Eq.~(\ref{self2}):
\bea      \label{approx1}
&&\int_{1}^{\infty} dx \,g(\Omega{+}K\Delta x)\,
\left(x-\sqrt{x^{2}-1}\right)
+\int_{-\infty}^{-1} dx \,g(\Omega{+}K\Delta x)\,
\left(x+\sqrt{x^{2}-1}\right)
\nonumber   \\
&\quad& = \int_{1}^{\infty} dx [g(\Omega{+}K\Delta x)
- g(\Omega{-}K\Delta x)] \left(x-\sqrt{x^{2}-1}\right)
\nonumber   \\
&\quad& \approx \int_{1}^{\infty} dx \,2\Omega \,g^{\prime}( K\Delta x)\,
\left(x-\sqrt{x^{2}-1}\right),
\eea
and expand Eq.~(\ref{self2}) to the order $(K\Delta)^3$. 
This yields  
\bea    \label{small}
&&\Delta \cos [(\Omega_{0}+\Omega_{1}(K\Delta)+\Omega_{2}(K\Delta)^2)\tau]
  \nonumber    \\
&&~= \frac{1}{2}\sqrt{\frac{\pi}{2}} e^{-{\Omega_{0}}^2/2}
  \left\{K\Delta-\Omega_{0}\Omega_{1}(K\Delta)^2
        + \left[({\Omega_{0}}^2-1)\left(\frac{{\Omega_{1}}^2}{2}
               +\frac{1}{8}\right)-\Omega_{0}\Omega_{2}\right](K\Delta)^3
   \right\}, \nonumber \\
&&\Delta \sin[(\Omega_{0}+\Omega_{1}(K\Delta)+\Omega_{2}(K\Delta)^2)\tau]
   \nonumber   \\
&&~= -\frac{\Omega_{0}}{2}K\Delta
    +\left[\frac{\Omega_{0}}{3}\sqrt{\frac{2}{\pi}}
           \left(1- e^{-{\Omega_{0}}^2/2}\right)
       -\frac{\Omega_{1}}{2}\right](K\Delta)^2
   \nonumber     \\
&&~ + \left\{\frac{\Omega_{0}}{8}+\frac{\Omega_{1}}{3}\sqrt{\frac{2}{\pi}}
         \left[ 1+({\Omega_{0}}^2-1)e^{-{\Omega_{0}}^2/2}\right]
       - \frac{\Omega_{2}}{2}\right\} (K\Delta)^3,
\eea
where Eqs.~(\ref{ao}) and the Gaussian distribution $g(\omega)$ have been used. After a
tedious calculation, we obtain from Eq.~(\ref{small})
\bea   \label{omega1}
\Omega_0\tau
&=& -\tan^{-1} \left(\sqrt{\frac{2}{\pi}}\Omega_0 e^{\Omega_0^2/2}\right),
   \nonumber   \\
\Omega_1\tau
&=& \sqrt{\frac{8}{\pi}}e^{\Omega_0^2/2} \left(1+\frac{2}{\pi}\Omega_0^2
 e^{\Omega_0^2}\right)^{-1} \left[\sqrt{\frac{2}{9\pi}}\Omega_0
     \left(1- e^{-\Omega_0^2/2} \right) -{\Omega_1\over2}- {\Omega_0^2\Omega_1\over2}
   \right], \nonumber  \\
\Omega_2\tau
&=& \sqrt{\frac{2}{\pi}}e^{-\Omega_0^2/2}\left(1+\frac{2}{\pi}\Omega_0^2
 e^{\Omega_0^2}\right)^{-1} \left\{(1+\Omega_0^2)\left({\Omega_0\over 8}
-\Omega_2\right) \right. \nonumber  \\
& &~ + \frac{8}{\pi} \Omega_0 e^{\Omega_0^2/2} 
        \left(1+\frac{2}{\pi}\Omega_0^2 e^{\Omega_0^2}\right)^{-1}
     \left[\sqrt{\frac{2}{9\pi}}\Omega_0 \left(1- e^{-\Omega_0^2/2} \right)
          -{\Omega_1\over2}- {\Omega_0^2\Omega_1\over2}\right]^2
  \nonumber  \\
& &~ - \left. {1\over 2}\Omega_0 \Omega_1^2 (\Omega_0^2+3) 
   + \sqrt{\frac{8}{9\pi}}\Omega_1 \left(1+ \Omega_0^2 - e^{-\Omega_0^2/2} \right) \right\}
\eea
together with
\be    \label{sc1}
\Delta = a_1 K\Delta+b_1 (K\Delta)^2+c_1 (K\Delta)^3+{\cal O}(K\Delta)^4,
\ee
which is just Eq.~(\ref{asc1}). The coefficients depend on $\Omega_0$, $\Omega_1$,
and $\Omega_2$ according to  
\bea
\label{cof1}
a_1 &=& \left(\frac{\Omega_0^2}{4}+\frac{\pi}{8} e^{-\Omega_0^2}\right)^{-1/2},
   \nonumber   \\
b_1 &=& \left(\Omega_0^2+{\pi\over2} e^{-\Omega_0^2}\right)^{-1/2}
        \left[-\sqrt{\frac{2}{9\pi}}\Omega_0^2 \left(1- e^{-\Omega_0^2/2}\right) 
         + \frac{\Omega_0\Omega_1}{2}\left(1-{\pi\over2} e^{-\Omega_0^2}\right)
        \right],
   \nonumber   \\
c_1 &=& \left(\Omega_0^2+{\pi\over2} e^{-\Omega_0^2}\right)^{-1/2}
      \left\{ \left({2\over9\pi}-{1\over8}\right)\Omega_0^2 + \frac{\Omega_0\Omega_1}{2}
      +{\Omega_1^2\over 4}-\sqrt{\frac{2}{9\pi}}\Omega_0\Omega_1 (2+\Omega_0^2)
       \right. \nonumber   \\
  && ~- \left(\Omega_0^2+{\pi\over2}  e^{-\Omega_0^2}\right)^{-1}
      \left[\sqrt{\frac{2}{9\pi}}\Omega_0^2 \left(1- e^{-\Omega_0^2/2}\right) 
       - \frac{\Omega_0\Omega_1}{2}\left(1-{\pi\over2} e^{-\Omega_0^2}\right)
      \right]^2
   \nonumber   \\
  && ~- \left({4\over 9\pi}\Omega_0^2-\sqrt{\frac{8}{9\pi}}\Omega_0 \Omega_1\right) e^{-\Omega_0^2/2}
   \nonumber  \\
  && ~\left.
    + \left[{2\over9\pi}\Omega_0^2+{\pi\over4}\left({\Omega_0^2\over8}
-{1\over8}+ \Omega_0^2\Omega_1^2 - {\Omega_1^2\over 2}-\Omega_0\Omega_2\right)\right]
   e^{-\Omega_0^2} \right\}.
\eea

\section*{Appendix B}
\renewcommand{\theequation}{B.\arabic{equation}}
\setcounter{equation}{0}

In the case $\Omega \gg K\Delta \,(\ll 1)$,
we approximate the integral in Eq.~(\ref{self2}) as follows:
\bea      \label{approx2}
&&\int_{1}^{\infty} dx g(\Omega{+}K\Delta x)\,
 \left(x-\sqrt{x^{2}-1}\right)
+\int_{-\infty}^{-1} dx \,g(\Omega{+}K\Delta x)\,
\left(x+\sqrt{x^{2}-1}\right)
\nonumber   \\
&\quad& = \int_{1}^{\infty} dx \,[g(\Omega{+}K\Delta x)
- g(\Omega{-}K\Delta x)] \left(x-\sqrt{x^{2}-1}\right)
\nonumber   \\
&\quad& \approx - \int_{1}^{\infty} dx \,g(\Omega{-}K\Delta x)\,
 \left(x-\sqrt{x^{2}-1}\right),
\eea
which, upon expansion to the order $(K\Delta)^3$, 
gives Eq.~(\ref{self2}) in the form 
\bea    \label{large}
&&\Delta \cos [(\Omega_{0}+\Omega_{1}(K\Delta)+\Omega_{2}(K\Delta)^2)\tau]
  \nonumber    \\
&&~~= \frac{1}{2}\sqrt{\frac{\pi}{2}} e^{-{\Omega_{0}}^2/2}\left\{K\Delta
-\Omega_{0}\Omega_{1}(K\Delta)^2
+ \left[({\Omega_{0}}^2-1)\left(\frac{{\Omega_{1}}^2}{2}
+\frac{1}{8}\right)-\Omega_{0}\Omega_{2}\right](K\Delta)^3
\right\}, \nonumber \\
&&\Delta \sin [(\Omega_{0}+\Omega_{1}(K\Delta)+\Omega_{2}(K\Delta)^2)\tau]
\nonumber   \\
&&~~= -\frac{(1+{\Omega_{0}}^2)}{4{\Omega_{0}}^3}
   \left[1+\Phi\left(\Omega_0/\sqrt{2}\right)\right]K\Delta
\nonumber  \\
&&~~+ \frac{1}{12{\Omega_{0}}^4}\left\{-\sqrt{\frac{2}{\pi}} e^{-{\Omega_{0}}^2/2}(4
{\Omega_{0}}^5+ 3{\Omega_{0}}^3+3\Omega_{0}\Omega_{1})
+ 3\Omega_{1}(3+{\Omega_{0}}^2)\left[1+\Phi\left(\Omega_0/\sqrt{2}\right)\right]
      \right\}(K\Delta)^2
\nonumber  \\
&&~~+ \frac{1}{48{\Omega_{0}}^5}
  \left\{\sqrt{\frac{2}{\pi}} e^{-{\Omega_{0}}^2/2}
        \left[16{\Omega_{0}}^7\Omega_{1}+2{\Omega_{0}}^5(3-14\Omega_{1}+3{\Omega_{1}}^2)
            -12{\Omega_{0}}^4\Omega_{2}
         \right.\right.
      \nonumber  \\
&&~~~~\left. +6{\Omega_{0}}^3(-1-2\Omega_{1}+3{\Omega_{1}}^2) +9\Omega_{0}(1+4{\Omega_{1}}^2)
        - 12{\Omega_{0}}^2\Omega_{2}\right]
        \nonumber \\
&&~~~~\left. -3\left[6+24{\Omega_{1}}^2 +{\Omega_{0}}^2(1+4{\Omega_{1}}^2)
       -12\Omega_{0}\Omega_{2}-4{\Omega_{0}}^3\Omega_{2}\right]
        \left[1+\Phi\left(\Omega_0/\sqrt{2}\right)\right]\right\}(K\Delta)^3
\eea
with the error function $\Phi(y)\equiv (2/\sqrt{\pi})\int_0^y dz~e^{-z^2}$.
After a tedious calculation, we obtain from Eq.~(\ref{large}):
\bea   \label{omega2}
\Omega_0\tau
&=& -\tan^{-1} \left\{ {1\over \sqrt{2\pi}} \left({1+\Omega_0^2\over\Omega_0^3}\right) 
        e^{\Omega_0^2/2}\left[1+ \Phi\left(\Omega_0/\sqrt{2}\right) \right]\right\} + \pi,
\nonumber   \\
\Omega_1\tau
&=& \left\{1+{1\over 2\pi}{(1+\Omega_0^2)^2 \over \Omega_0^6} e^{\Omega_0^2}
          \left[1+ \Phi\left(\Omega_0/\sqrt{2}\right)\right]^2 \right\}^{-1} \nonumber \\
   & &~~~\times \left\{-{4\Omega_0\over 3\pi} -{\Omega_1-1\over\pi\Omega_0}- {\Omega_1\over \pi\Omega_0^3}
         + {1\over \sqrt{2\pi}}{\Omega_1\over \Omega_0^4}
      (3-\Omega_0^4) e^{\Omega_0^2/2} \left[1+ \Phi\left(\Omega_0/\sqrt{2}\right)\right]  \right\},
   \nonumber  \\
\Omega_2\tau
&=& \left\{ {1\over \sqrt{2\pi}} \left({1+\Omega_0^2\over\Omega_0^3}\right) 
           e^{\Omega_0^2/2} \left[1+ \Phi\left(\Omega_0/\sqrt{2}\right)\right] 
    \right\}
    \left\{1+{1\over 2\pi}{(1+\Omega_0^2)^2 \over \Omega_0^6} e^{\Omega_0^2}
      \left[1+ \Phi\left(\Omega_0/\sqrt{2}\right)\right]^2 \right\}^{-2}
     \nonumber  \\
& &~~~\times \left\{-{4\Omega_0\over 3\pi} -{\Omega_1-1\over\pi\Omega_0}
     - {\Omega_1\over \pi\Omega_0^3}
  + {1\over \sqrt{2\pi}}{\Omega_1\over \Omega_0^4}
     (3-\Omega_0^4) e^{\Omega_0^2/2}\left[1+ \Phi\left(\Omega_0/\sqrt{2}\right)\right] \right\}
   \nonumber  \\
& &~~- \left\{ {1\over\sqrt{2\pi}\Omega_0^3} \left[{1\over8}
     (1-\Omega_0^4)-{\Omega_1^2\over2}(5-\Omega_0^4)+\Omega_0\Omega_2(1+\Omega_0^2)\right] 
     e^{\Omega_0^2/2} \left[1+ \Phi\left(\Omega_0/\sqrt{2}\right)\right] \right.
 \nonumber  \\
& &~~+ {1\over 12\pi\Omega_0^4}[16\Omega_0^6\Omega_1+2\Omega_0^4(3-14\Omega_1
   +3\Omega_1^2)-12\Omega_0^3\Omega_2-6\Omega_0^2(1+2\Omega_1-3\Omega_1^2)
 \nonumber  \\
& &~~ -12 \Omega_0\Omega_2+9(1+4\Omega_1^2)]-{1\over 4\sqrt{2\pi}\Omega_0^5}
  [6+24\Omega_1^2+\Omega_0^2(1+4\Omega_1^2)-12\Omega_0\Omega_2-4\Omega_0^3 \Omega_2]
 \nonumber   \\
& &~~~~\left.\times e^{\Omega_0^2/2} \left[1+ \Phi\left(\Omega_0/\sqrt{2}\right)\right]
 \right\}
\eea
and
\be    \label{sc2}
\Delta = a_2 K\Delta+b_2 (K\Delta)^2+c_2 (K\Delta)^3+{\cal O}(K\Delta)^4,
\ee
which is Eq.~(\ref{asc2}).
Again the coefficients depend on $\Omega_0$, $\Omega_1$, and $\Omega_2$ via
\bea   \label{cof2}
a_2
&=& \left\{\frac{\pi}{8} e^{-\Omega_0^2}+{(1+\Omega_0^2)^2\over 16\Omega_0^6}
\left[1+ \Phi\left(\Omega_0/\sqrt{2}\right)\right]^2\right\}^{-1/2},
\nonumber   \\
b_2
&=& \left\{\frac{\pi}{2} e^{-\Omega_0^2}+{(1+\Omega_0^2)^2\over 4\Omega_0^6}
 \left[1+ \Phi\left(\Omega_0/\sqrt{2}\right)\right]^2 \right\}^{-1/2}
 \left\{ -{\pi\over4}\Omega_0\Omega_1 e^{-\Omega_0^2}+{1+\Omega_0^2\over 8\Omega_0^6}
       \left[1+ \Phi\left(\Omega_0/\sqrt{2}\right)\right] \right.
     \nonumber  \\
&&~~~\left.\times\left[{2\over\pi}e^{-\Omega_0^2/2}
                  \left({4\over3}\Omega_0^4+\Omega_0^2(\Omega_1-1)+\Omega_1\right)
                -(3+\Omega_0^2)\Omega_1\left[1+ \Phi\left(\Omega_0/\sqrt{2}\right)\right]\right]
     \right\},  \nonumber   \\
c_2
&=& \left\{\frac{\pi}{2} e^{-\Omega_0^2}+{(1+\Omega_0^2)^2\over 4\Omega_0^6}
      \left[1+ \Phi\left(\Omega_0/\sqrt{2}\right)\right]^2 \right\}^{-3/2}
    \left\{-{\pi\over4}\Omega_0\Omega_1 e^{-\Omega_0^2}+{1+\Omega_0^2\over 8 \Omega_0^6}
            \left[1+ \Phi\left(\Omega_0/\sqrt{2}\right)\right] \right.
      \nonumber  \\
&&~~~\left.\times\left[\sqrt{\frac{2}{\pi}}e^{-\Omega_0^2/2}
                 \left({4\over3}\Omega_0^4+\Omega_0^2(\Omega_1-1)+\Omega_1\right)
                 -(3+\Omega_0^2)\Omega_1\left[1+ \Phi\left(\Omega_0/\sqrt{2}\right)\right]
           \right]
    \right\}^2  \nonumber  \\
&&~~+ \left\{\frac{\pi}{2} e^{-\Omega_0^2}+{(1+\Omega_0^2)^2\over 4\Omega_0^6}
             \left[1+ \Phi\left(\Omega_0/\sqrt{2}\right)\right]^2 \right\}^{-1/2}
     \left\{ {\pi\over 8} e^{-\Omega_0^2}\left(2\Omega_0^2\Omega_1^2
                     +{\Omega_0^2\over4}-\Omega_1^2-2\Omega_0\Omega_2-{1\over4}\right)
     \right. \nonumber   \\
&&~~+ {1\over 144\Omega_0^7}\left[\sqrt{\frac{2}{\pi}}e^{-\Omega_0^2/2}
           \left({4\over3}\Omega_0^4+\Omega_0^2(\Omega_1-1)+\Omega_1\right)
           -(3+\Omega_0^2)\Omega_1\left[1+ \Phi\left(\Omega_0/\sqrt{2}\right)\right]\right]^2
  \nonumber  \\
&&~~-\sqrt{\frac{2}{\pi}}e^{-\Omega_0^2/2}{1+\Omega_0^2\over 96 \Omega_0^7}
      \left[16\Omega_0^6\Omega_1+2\Omega_0^4(3-14\Omega_1+3\Omega_1^2)-12\Omega_0^3\Omega_2 \right.
    \nonumber   \\
&&~~\left.-6\Omega_0^2(1+2\Omega_1-3\Omega_1^2)-12 \Omega_0\Omega_2+9(1+4\Omega_1^2)\right]
     \left[1+ \Phi\left(\Omega_0/\sqrt{2}\right)\right]
     \nonumber \\
&&~~\left.-{1\over 4\sqrt{2\pi}\Omega_0^5}
     \left[6+24\Omega_1^2+\Omega_0^2(1+4\Omega_1^2)-12\Omega_0\Omega_2-4\Omega_0^3\Omega_2\right]
     \right\}.
\eea

\newpage


\begin{figure}[H]
\begin{tabular}{cccc}
\epsfig{width=13cm,height=7cm,file=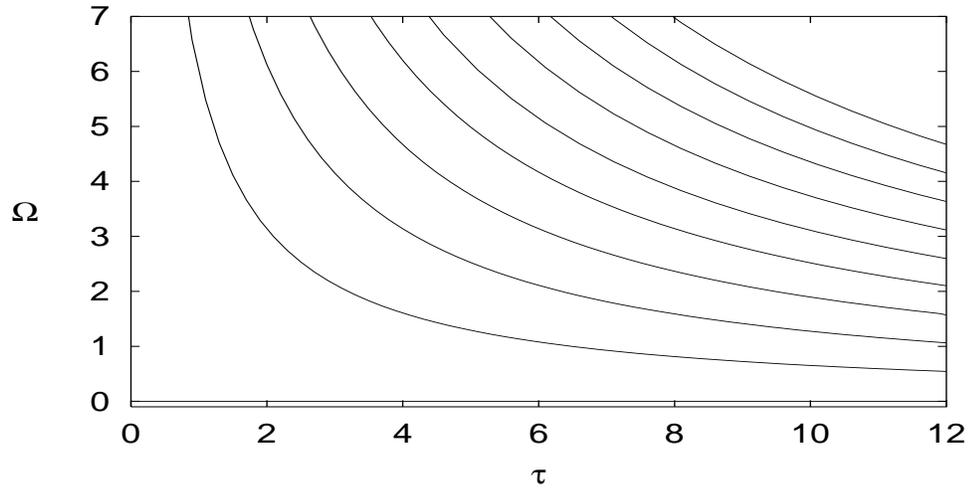}    \\
\hspace{1cm} $(a)$ \vspace{1cm}     \\
\epsfig{width=13cm,height=7cm,file=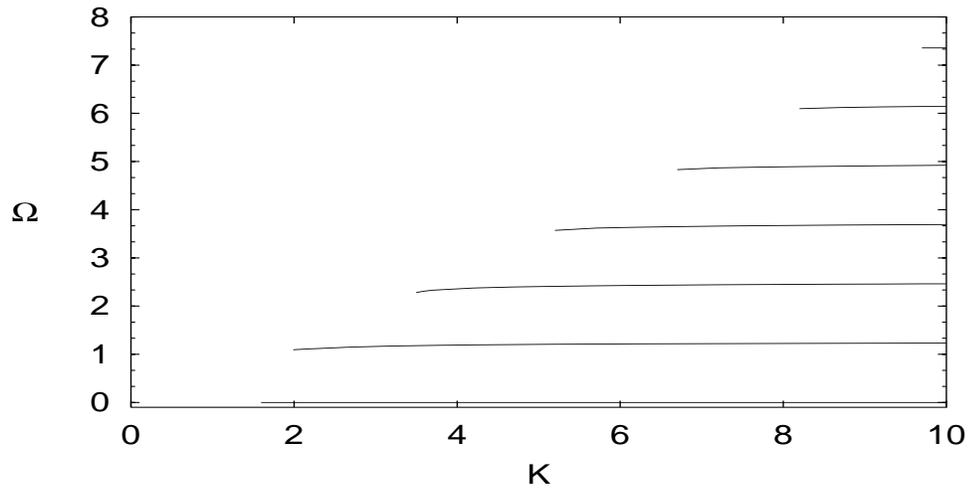}    \\
\hspace{1cm} $(b)$  \vspace{1cm}    \\
\end{tabular}
\caption[Synchronization frequency as functions of the delay time and
coupling strength.]{Synchronization frequency $\Omega$
(a) as functions of the delay time $\tau$ at the coupling strength $K = 10$,
where frequency suppression and multistability can be observed;
(b) as functions of $K$ at $\tau = 5$.
For given $K$,
the largest synchronization frequency belonging to the highest stair 
gives the most stable solution, but the corresponding basin of attraction
is small.}
\end{figure}

\begin{figure}[H]
\begin{tabular}{cccc}
\epsfig{width=13cm,height=7cm,file=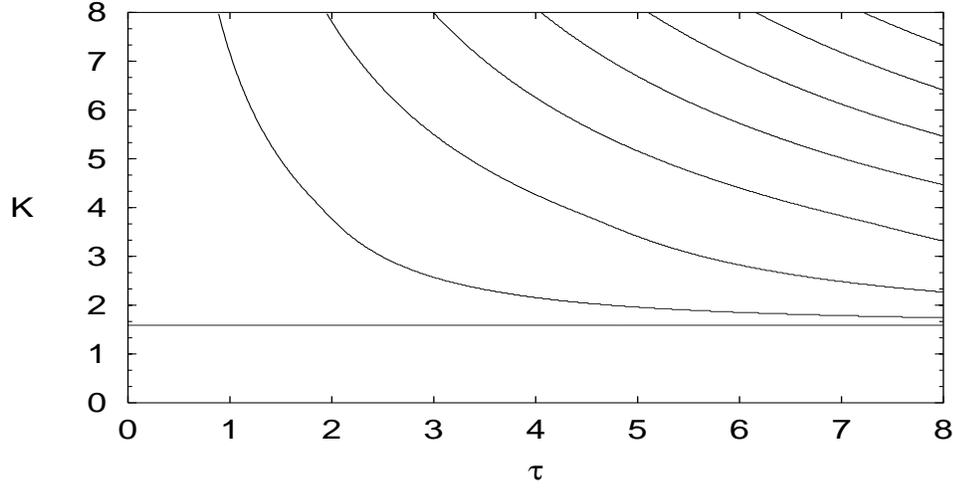}    \\
\hspace{1cm} $(a)$ \vspace{0.5cm}     \\
\epsfig{width=13cm,height=7cm,file=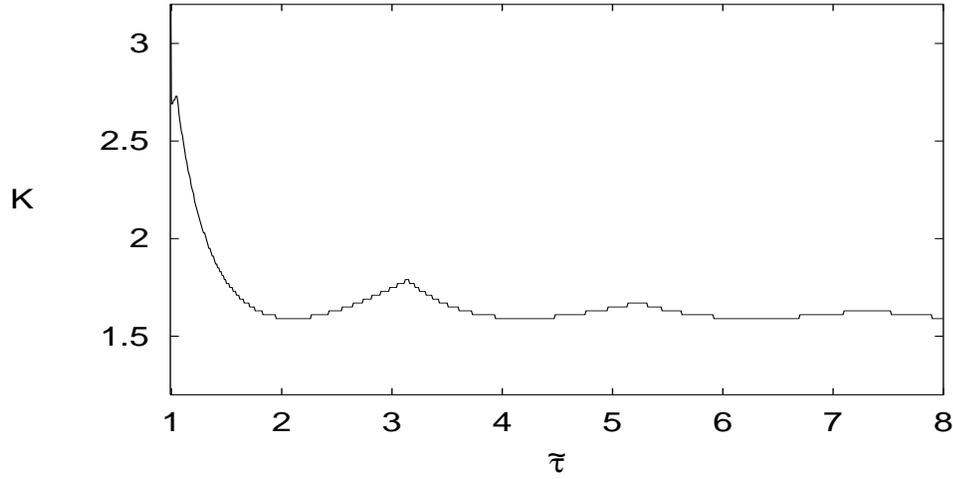}    \\
\hspace{1cm} $(b)$  \vspace{0.5cm}    \\
\end{tabular}
\caption[Phase boundaries between the incoherent and coherent states.]
{Phase diagram on the $K$-$\tau$ plane, displaying boundaries between the incoherent 
and coherent states.  
Whenever each boundary in (a) is crossed from below,
a new additional coherent state with a larger synchronization frequency emerges.
Below the lowest boundary, which is the straight line $K=K_c \approx 1.596$, 
only the incoherent state exists.
In (b) the boundaries are redrawn, with the horizontal axis rescaled:
$\tilde{\tau}$ in (b) corresponds to $\Omega\tau/(\Omega+3)$ in (a).
Here displayed is only the envelope consisting of the curve segments with lowest values of $K$ 
at given $\tilde{\tau}$, which describes the phase boundary
separating the incoherent state in the system with $\omega_0 =3$.
}
\end{figure}
                                  
\begin{figure}[H]
\centerline{
\epsfig{width=13cm,height=7cm,file=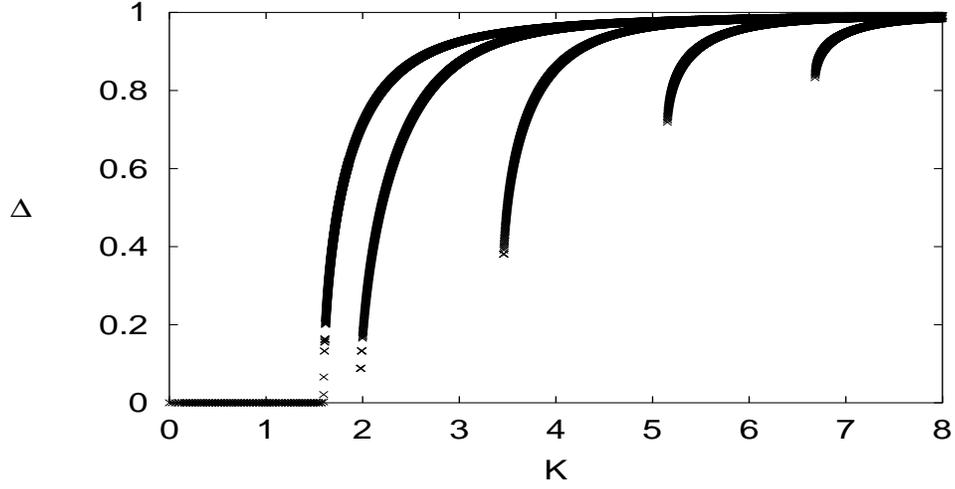}}
\vspace{1cm}
\caption[Order parameter as a function of the coupling strength.]
{Order parameter $\Delta$ as a function of the coupling strength $K$ at
the time delay $\tau = 5$.
Each line describes the transition for each synchronization
frequency, and the critical coupling strength $K_c$ is shown to increase for large
$\Omega$. The leftmost curve, which corresponds to $\Omega = 0$,
gives $K_c \approx 1.596$; the next one, corresponding to $\Omega
\approx 1.09$, gives $K_c \approx 1.97$. The numerical results displayed by the
leftmost curve agree well with the analytical ones.}
\end{figure}

\begin{figure}[H]
\centerline{
\epsfig{width=13cm,height=7cm,file=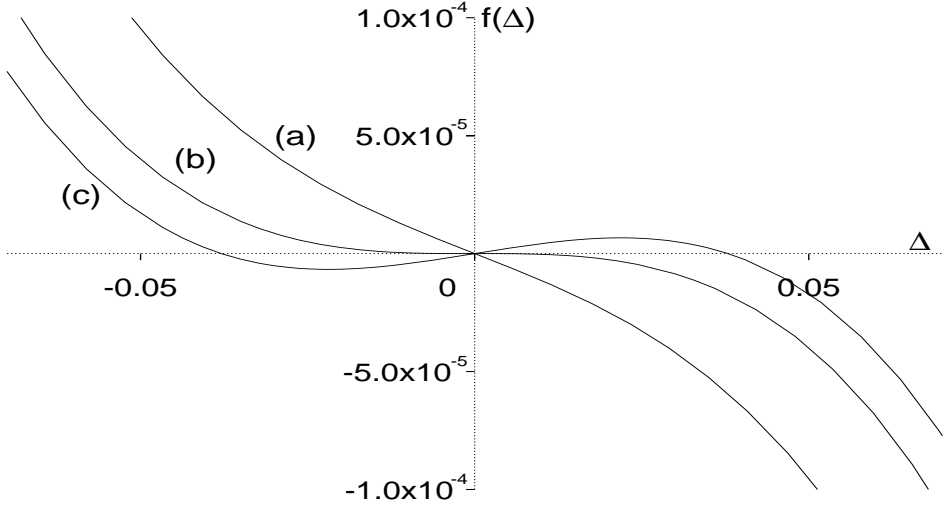}}
\vspace{1cm}
\caption{Graphical solutions of Eq.~(\ref{asc1}), displaying 
$f(\Delta)\equiv (a_1 K-1)\Delta +b_1 (K\Delta)^2+c_1 (K\Delta)^3$ 
versus $\Delta$, for $b_1 =0$ with
(a) $K=1.594 \,(<K_c)$, (b) $K=1.596 \,(=K_c)$, and (c) $K=1.597 \,(>K_c)$.
The negative solution appearing in (c) is unphysical.}
\end{figure}

\begin{figure}[H]
\begin{tabular}{cccc}
\epsfig{width=8cm,height=8cm,file=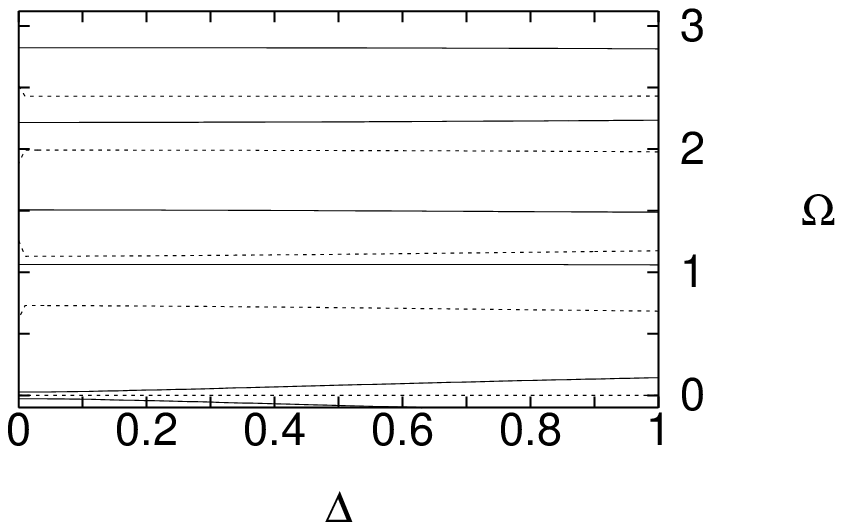} &
\epsfig{width=8cm,height=8cm,file=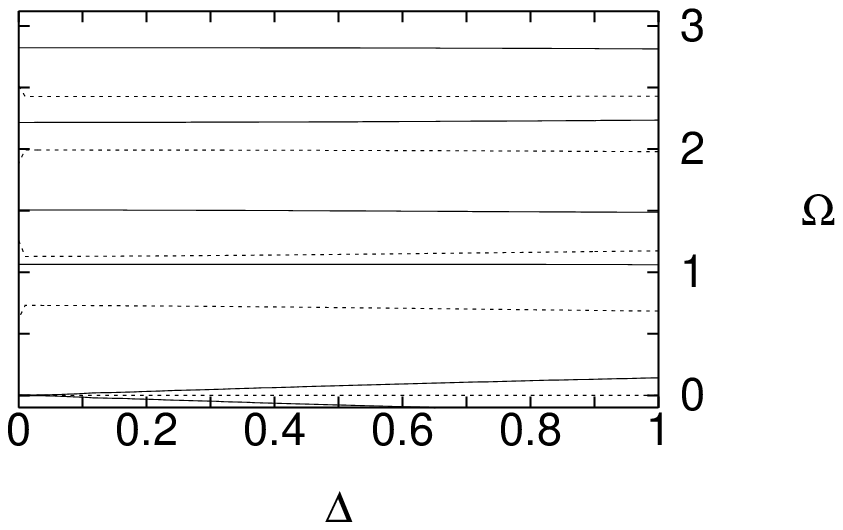} \vspace{-0.5cm} \\
$(a)$   &    $(b)$        \\
\epsfig{width=8cm,height=8cm,file=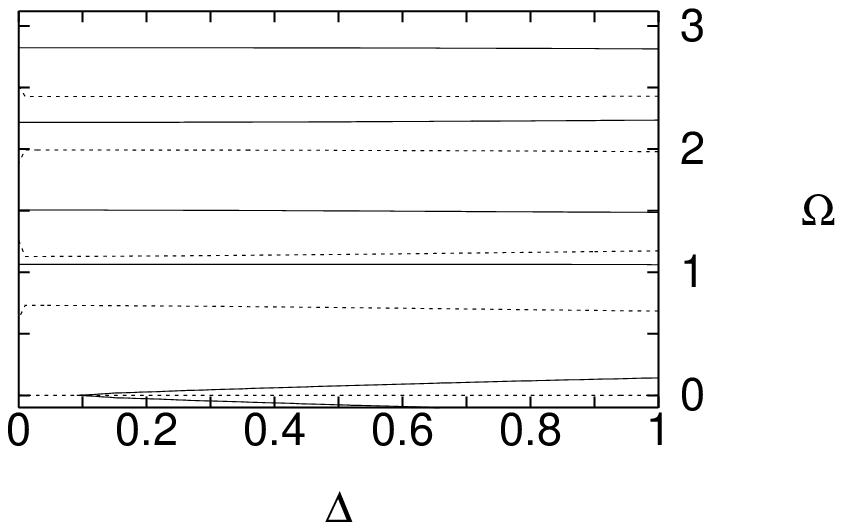}   &
\epsfig{width=8cm,height=8cm,file=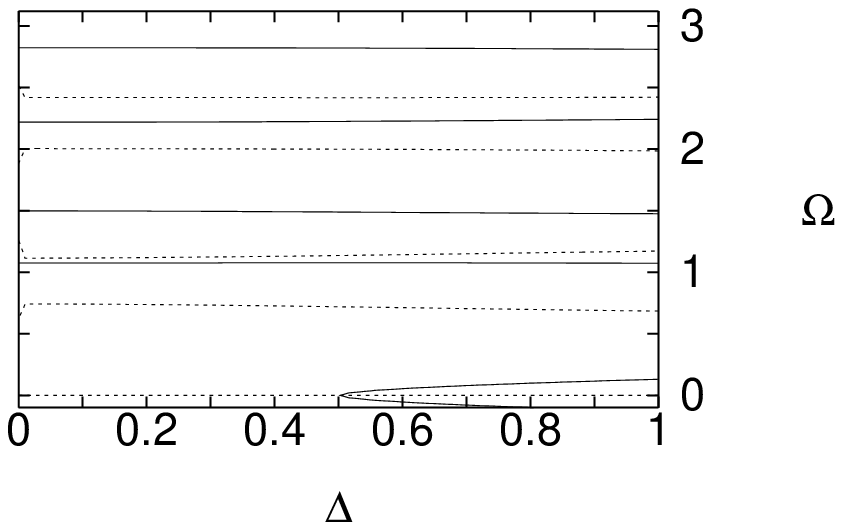} \vspace{-0.5cm} \\
$(c)$   &    $(d)$     \vspace{1cm}   \\
\end{tabular}
\caption[Synchronization frequency versus the order parameter, displaying
a continuous phase transition at $K = K_c \approx 1.595$.]
{Synchronization frequency $\Omega$ versus the order parameter $\Delta$
obtained from Eq.~(\ref{self2}) for $\tau=5$ at (a) $K=1.58$, (b) $K=1.596$,
(c) $K=1.60$, (d) $K=1.75$. Solid and dotted lines represent
solutions of the first equation and the second equation of Eq.~(\ref{self2}).
A continuous transition for $\Omega = 0$ can be observed at $K = 1.596$.}
\end{figure}

\begin{figure}[H]
\begin{tabular}{cccc}
\epsfig{width=8cm,height=8cm,file=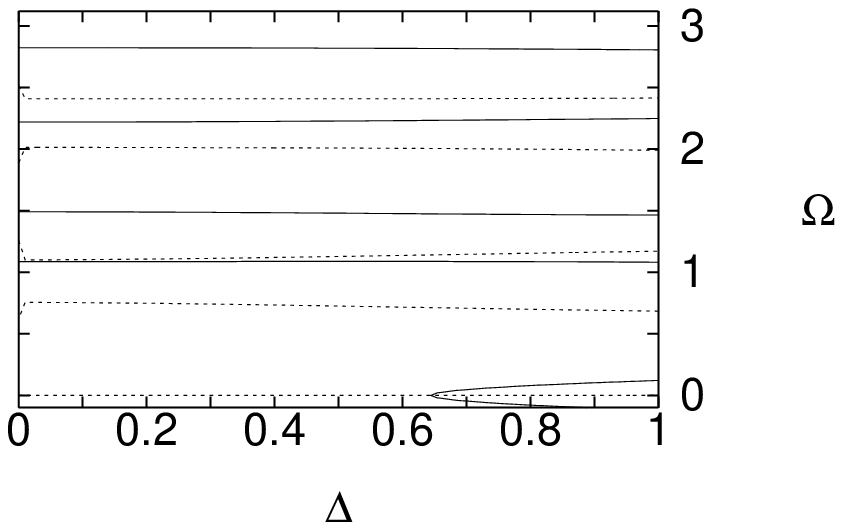} &
\epsfig{width=8cm,height=8cm,file=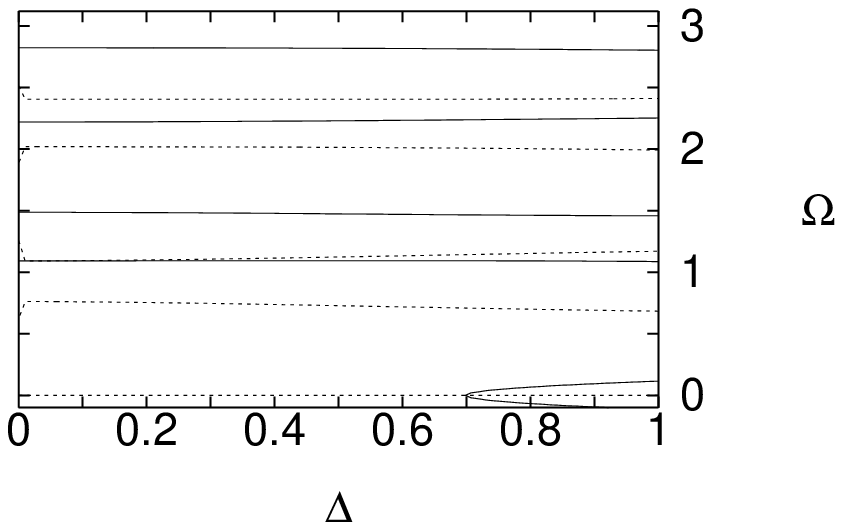} \vspace{-0.5cm} \\
$(a)$   &    $(b)$        \\
\epsfig{width=8cm,height=8cm,file=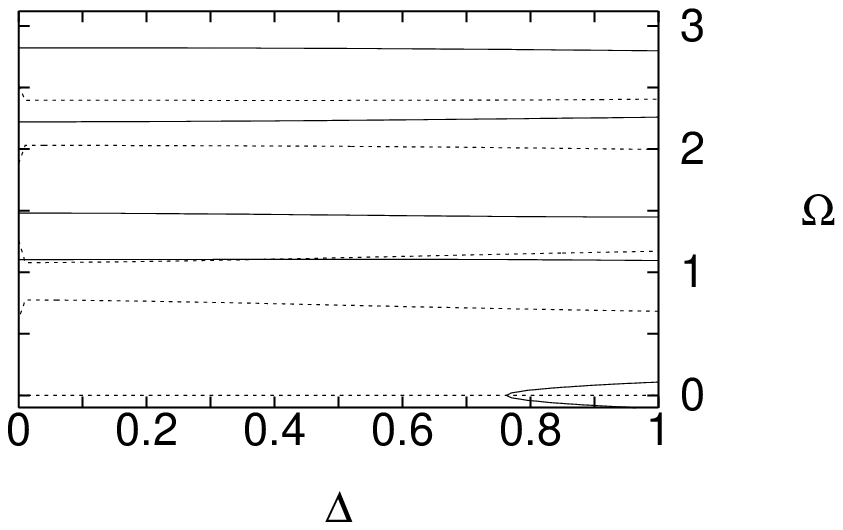}   &
\epsfig{width=8cm,height=8cm,file=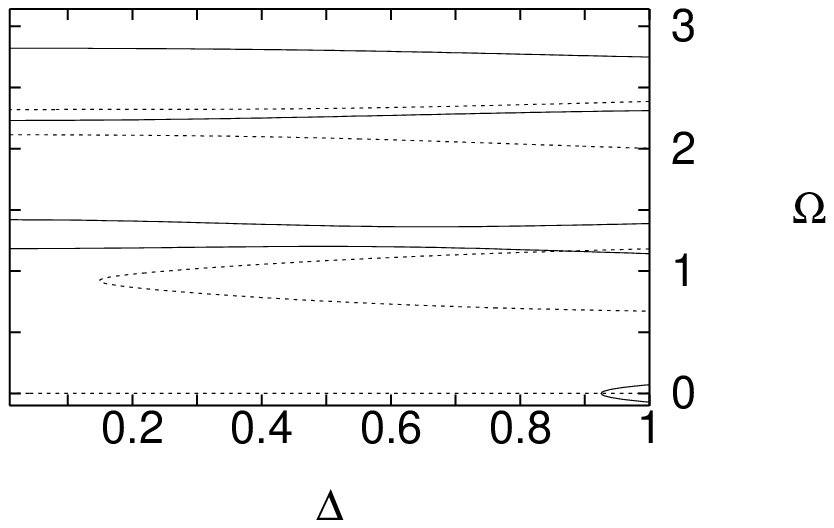} \vspace{-0.5cm} \\
$(c)$   &    $(d)$     \vspace{1cm}   \\
\end{tabular}
\caption[Synchronization frequency versus the order parameter, displaying
a discontinuous phase transition at $K \approx 1.97$.]
{Synchronization frequency $\Omega$ versus the order parameter $\Delta$
obtained from Eq.~(\ref{self2}) for $\tau=5$ at (a) $K=1.89$, (b) $K=1.97$,
(c) $K=2.10$, (d) $K=3.00$. 
A discontinuous transition for $\Omega \approx 1.09$ can be observed at $K = 1.97
$.}
\end{figure}

\begin{figure}[H]
\begin{tabular}{cccccc}
\epsfig{width=8cm,height=8cm,file=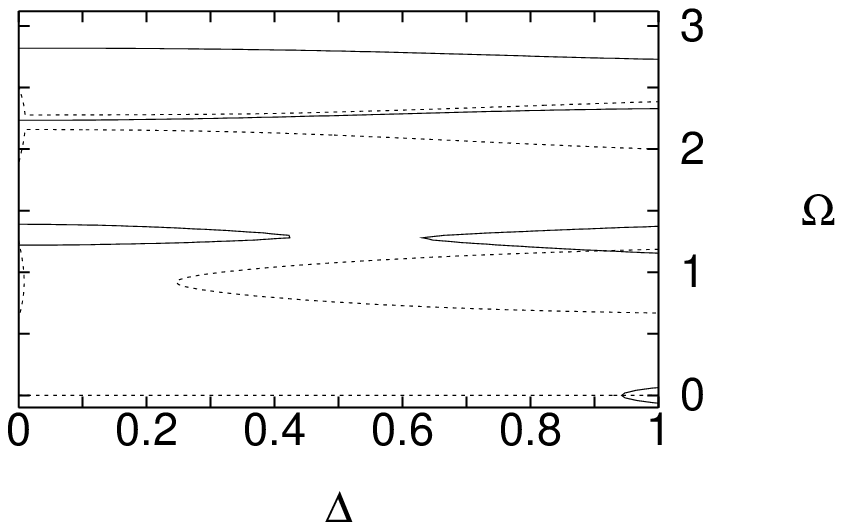} &
\epsfig{width=8cm,height=8cm,file=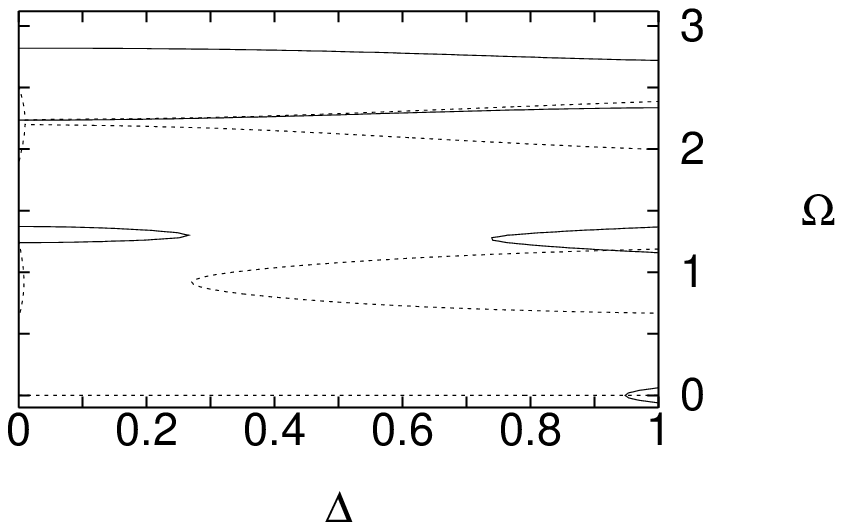} \vspace{-0.5cm} \\
$(a)$   &    $(b)$        \\
\epsfig{width=8cm,height=8cm,file=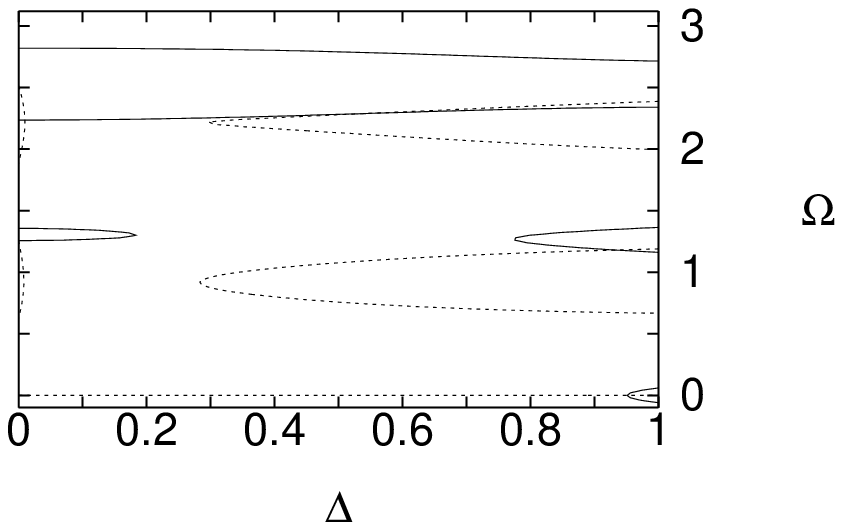}   &
\epsfig{width=8cm,height=8cm,file=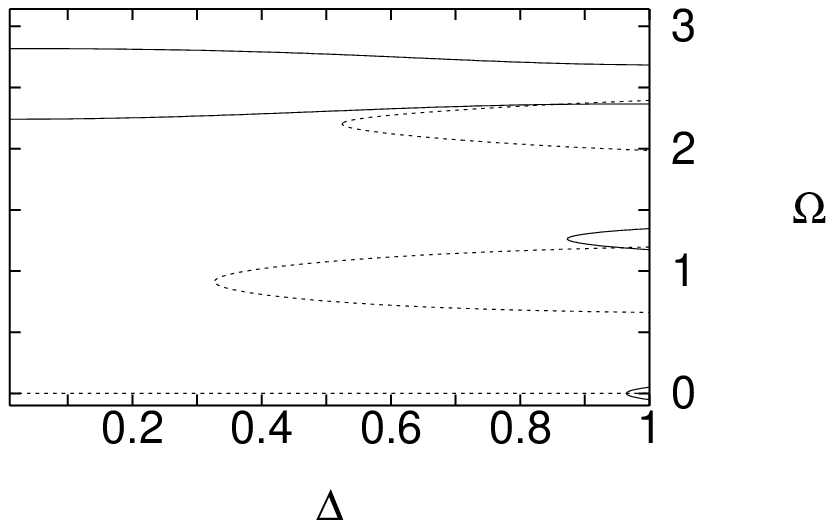} \vspace{-0.5cm} \\
$(c)$   &    $(d)$   \vspace{1cm}     \\
\end{tabular}
\caption[Synchronization frequency versus the order parameter, displaying
a discontinuous phase transition at $K \approx 3.46$.]
{Synchronization frequency $\Omega$ versus the order parameter $\Delta$
obtained from Eq.~(\ref{self2}) for $\tau=5$ at (a) $K=3.30$, (b) $K=3.46$,
(c) $K=3.515$, (d) $K=4.00$. 
A discontinuous transition for $\Omega \approx 2.25$ can be observed at $K = 3.46$.
}
\end{figure}

%
\begin{figure}[H]
\begin{tabular}{cc}
\epsfig{width=10cm,height=6cm,file=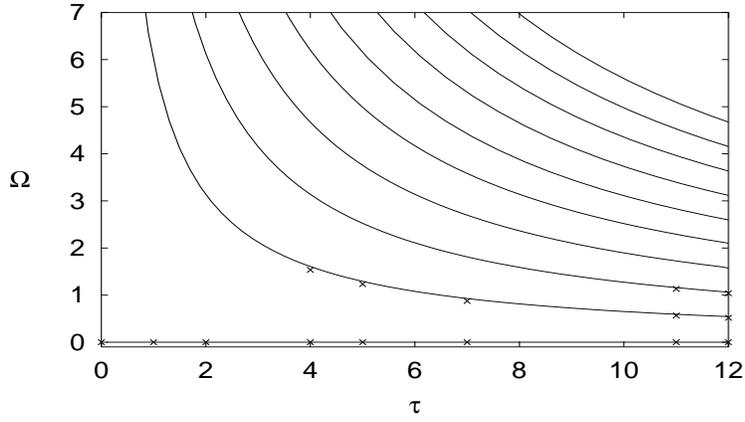}    \\
\hspace{1cm} $(a)$ \vspace{1cm}     \\
\epsfig{width=10cm,height=6cm,file=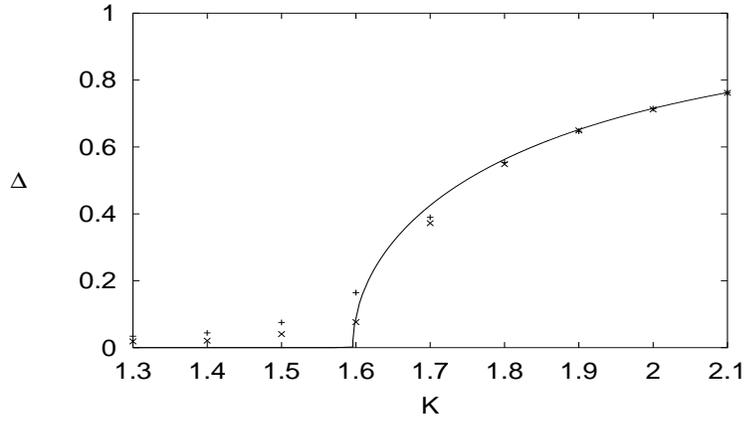}    \\
\hspace{1cm} $(b)$ \vspace{1cm}     \\
\end{tabular}
\caption[Results of numerical simulations.]
{Results of numerical simulations on 5000 coupled oscillators:
(a) Synchronization frequency versus the delay time.
Crosses are results of numerical simulations and
solid lines represent the solutions of Eq.~(\ref{self2}),
displaying perfect agreement with each other.
(b) Order parameter versus the coupling strength, displaying continuous
transitions (with zero synchronization frequency) for $\tau = 0$ (plus signs)
and $\tau = 5$ (crosses).
The solid line represents analytic results for $\tau=5$, 
displaying reasonable agreement with the numerical ones.
Slight suppression of synchronization by time delay can be observed
near the transition region.
The size of the error bars estimated by the standard deviation is about 
that of the symbols and lines are merely guides to the eye.}
\end{figure}


\end{document}